\documentclass[11pt,a4paper]{article}
\usepackage{jheppub}

\usepackage{graphicx,epsfig}
\usepackage{amsmath}
\usepackage[normalem]{ulem} 
\usepackage{amsfonts} 
\usepackage{amssymb}

\newcommand{\bea}{\begin{eqnarray}}
\newcommand{\beq}{\begin{equation}}
\newcommand{\eea}{\end{eqnarray}}
\newcommand{\eeq}{\end{equation}}
\newcommand\brabar{\raisebox{-4.0pt}{\scalebox{.2}{ \textbf{(}}}\raisebox{-4.0pt}{{\_}}\raisebox{-4.0pt}{\scalebox{.2}{\textbf{)}}}}

\newcommand{\jhep}{\emph{JHEP }}
\newcommand{\jcap}{\emph{JCAP }}
\newcommand{\prd}{\emph{Phys.\ Rev.} {\bf D }}
\newcommand{\prl}{\emph{Phys.\ Rev.\ Lett. }}
\newcommand{\plb}{\emph{Phys.\ Lett.} {\bf B }}
\newcommand{\app}{\emph{Astropart.\ Phys. }}

\newcommand{\mpla}{\emph{Mod.\ Phys.\ Lett.} {\bf A }}

\newcommand{\npps}{\emph{Nucl.\ Phys.\ Proc.\ Suppl. }}

\newcommand{\ptp}{\emph{Prog.\ Theor.\ Phys. }}
\newcommand{\npb}{\emph{Nucl.\ Phys.} {\bf B }}
\newcommand{\zpc}{\emph{Z.\ Phys.} {\bf C }}
\newcommand{\ap}{\emph{Annals Phys. }}

\newcommand{\nima}{\emph{Nucl.\ Instrum.\ Meth.} {\bf A }}
\newcommand{\rpp}{\emph{Rept.\ Prog.\ Phys. }}
\newcommand{\ibid}{\emph{ibid }}

\title{SUSY renormalization group effects in ultra high energy neutrinos}

\author[a,b]{M.~Bustamante,}
\author[a]{A.M.~Gago,}
\author[c,d]{and J.~Jones P\'erez}

\affiliation[a]
{Secci\'on F\'isica, Departamento de Ciencias, 
Pontificia Universidad Cat\'olica del Per\'u, \\
Apartado 1761, Lima, Peru}
\affiliation[b]
{Theoretical Physics Department, Fermi National Accelerator Laboratory,\\
Batavia, IL 60510, USA}
\affiliation[c]
{Departament de F\'{\i}sica Te\`orica and IFIC, Universitat de Val\`encia-CSIC, \\
E-46100, Burjassot, Spain}
\affiliation[d]
{INFN, Laboratori Nazionali di Frascati, \\
Via E.~Fermi 40, I-00044 Frascati, Italy}

\emailAdd{mbustamante@pucp.edu.pe}
\emailAdd{agago@pucp.edu.pe}
\emailAdd{joel.jones@lnf.infn.it}

\abstract{
We have explored the question of whether the renormalization group running of the neutrino mixing parameters in the Minimal Supersymmetric Standard Model is detectable with ultra-high energy neutrinos from active galactic nuclei (AGN).  We use as observables the ratios of neutrino fluxes produced at the AGN, focusing on four different neutrino production models: $\left(\Phi_{\nu_e+\bar{\nu}_e}^0 : \Phi_{\nu_\mu+\bar{\nu}_\mu}^0 : \Phi_{\nu_\tau+\bar{\nu}_\tau}^0\right)$ = $(1:2:0)$, $(0:1:0)$, $(1:0:0)$, and $(1:1:0)$. The prospects for observing deviations experimentally are taken into consideration, and we find out that it is necessary to impose a cut-off on the transferred momentum of $Q^2 \geq 10^7$ GeV$^2$. However, this condition, together with the expected low value of the diffuse AGN neutrino flux, yields a negligible event rate at a km-scale \u{C}erenkov detector such as IceCube.
}


\arxivnumber{1012.2728}

\begin{document}
 
\maketitle

\section{Introduction}

Experiments performed over the last ten years have confirmed the mass-induced neutrino oscillation phenomenon, caused by the non-coincidence of mass and flavour eigenstates. Rather, the flavour states, $\nu_e$, $\nu_\mu$, and $\nu_\tau$, are linear superpositions of the mass eigenstates, which are customarily denoted by $\nu_1$, $\nu_2$, and $\nu_3$, i.e., 
\begin{equation}
 \nu_\alpha = \sum_i U_{\alpha i}^\ast \nu_i ~,
\end{equation}
with $\alpha = e, \mu, \tau$ and $i = 1,2,3$. The coefficients $U_{\alpha i}$ are components of the lepton mixing matrix, $U_\nu$, also known as the Pontecorvo-Maki-Nakagawa-Sakata (PMNS) matrix, that connects the bases of mass and flavour neutrinos and which can be parametrised in terms of three mixing angles, $\theta_{12}$, $\theta_{23}$ and $\theta_{13}$, and one CP-violating phase, $\delta$, as
\begin{eqnarray}\label{EqUPMNS}
 U_\nu
 = \left(\begin{array}{ccc}
     c_{12}c_{13}                                & s_{12}c_{13}                                & s_{13}e^{-i\delta} \\
     -s_{12}c_{23}-c_{12}s_{23}s_{13}e^{i\delta} & c_{12}c_{23}-s_{12}s_{23}s_{13}e^{i\delta}  & s_{23}c_{13} \\
     s_{12}s_{23}-c_{12}c_{23}s_{13}e^{i\delta}  & -c_{12}s_{23}-s_{12}c_{23}s_{13}e^{i\delta} & c_{23}c_{13}
 \end{array}\right) ,
\end{eqnarray}
where $c_{ij} \equiv \cos\left(\theta_{ij}\right)$ and $s_{ij} \equiv \sin\left(\theta_{ij}\right)$. A recent global analysis \cite{GonzalezGarcia:2010er} used data from solar (SNO, Chlorine, Gallex/GNO, Borexino, SAGE), reactor (KamLAND, CHOOZ), accelerator (K2K, MINOS), and atmospheric (Super-Kamiokande) neutrino experiments to find the following $3\sigma$ bounds on the mixing angles and mass differences:
\begin{equation}
 \Delta m_{21}^2 = 7.59^{+0.61}_{-0.69} \times 10^{-5} ~\text{eV}^2  ~, ~~ 
 \Delta m_{31}^2 =
 \left\{\begin{array}{l}
  \left(-2.36 \pm 0.37\right) \times 10^{-3} ~\text{eV}^2 \\
  \left(+2.46 \pm 0.37\right) \times 10^{-3} ~\text{eV}^2
 \end{array}\right. ~,
\label{masses}
\end{equation}
\begin{equation}
 \theta_{12} = 34.4\left.^{+3.2}_{-2.9}\right.^\circ ~, ~~
 \theta_{23} = 42.8\left.^{+10.7}_{-7.3}\right.^\circ ~, ~~
 \theta_{13} = 5.6\left.^{+6.9}_{-5.6}\right.^\circ ~, ~~
 \delta \in [0,2\pi] ~.
\label{mixing_angles}
\end{equation}
Note, however, that the experiments performed so far have probed neutrino energies from the few MeV (solar neutrinos) to the GeV range (atmospheric neutrinos). Thus, it is important to point out that the values of the mixing angles and mass differences found in this and similar analyses are valid, in principle, only in that energy range.

Ever since the oscillation phenomenon was first observed, neutrino physics has been considered a window towards physics beyond the Standard Model (SM) \cite{Mohapatra:2005wg}. Thus, the study of the influence of new physics on neutrino properties is of high importance. In the following, we shall address how loop corrections on neutrino interactions are affected by new physics, in particular that of the Minimal Supersymmetric Standard Model (MSSM).

In neutrino experiments, loop corrections can play a role at two sources. The first source comes from the neutrino interaction vertex, where diagrams with intermediate particles can modify the vertex by either introducing a correction due to form factors~\cite{Aoki:1980ix} or by changing the flavour structure of the mixing matrix entering the vertex itself (an example for the CKM matrix in the MSSM can be found in~\cite{Buras:2002vd}). At one loop, most of the corrections in the SM are proportional to the PMNS matrix, so we do not expect to obtain any significant corrections to the flavour structure. On the other hand, supersymmetric models like the MSSM introduce chargino, neutralino and slepton mixing matrices into the game, meaning that they have the potential to modify the flavour structure of the effective vertex.

As the vertex corrections need to be renormalized, an arbitrary scale dependence is introduced. This causes the well-known running of parameters through renormalization group equations (RGEs). In order to avoid large logarithms in the loop functions, it is customary to set the scale $\mu$ such that these large logarithms vanish. For vertex corrections, this means that the scale is of the order of the transferred momentum, $\mu\approx Q \equiv \sqrt{-q^2}$~\cite{Collins:1984xc}.

The evaluation of the Green function related to the vertex requires all parameters involved in the function to be set at the same scale. This implies that one needs to take into account the RGE evolution of all parameters according to the transferred momentum of the interaction. Moreover, as the RGE evolution of the parameters is independent of the process one is analysing, this constitutes an independent second source of corrections to the vertex. The running of the mixing angles has been thoroughly studied, for example, by~\cite{Babu:1993qv,Antusch:2001ck,Antusch:2001vn,Antusch:2003kp,Antusch:2005gp}. The difference in the evolution of the mixing parameters between the SM and MSSM was analysed in detail in~\cite{Antusch:2003kp,Antusch:2005gp}, where it was shown that if the bounds on neutrino masses are respected, the SM running is negligible. Thus, even though in this work we shall perform the SM running, we shall refer to it as the ``no-running'' scenario. The MSSM scenario, on the other hand, could provide an additional enhancement that allowed large running effects to take place.

Motivated by the results in~\cite{Antusch:2003kp}, in this work we study whether there could be any observable deviations from the SM neutrino oscillation framework probed at a very high energy scale, due only to the RGEs in the MSSM. To this end, in Section~\ref{sec:running} we briefly review the origin and structure of the RGE corrections. In Section~\ref{sec:setup} we establish how these corrections should be introduced within a neutrino experiment. In Section~\ref{sec:theory} we describe the changes one could expect if the RGE corrections are implemented from ultra high energy neutrinos produced by active galactic nuclei and examine the experimental requirements in order to observe the deviations. We conclude in Section~\ref{conclusion}.

\section{Running of Neutrino Mixing Parameters}
\label{sec:running}

In a SM with massive neutrinos, the RGE evolution of neutrino masses and mixing angles is a consequence of the renormalization of the effective neutrino dimension-five operator
\begin{equation}
 \label{dim5}
 \mathcal{L}_\nu=\frac{1}{4}(\overline{L}^c_i H)\frac{m^\nu_{ij}}{\Lambda_\nu}(L_j H) ~,
\end{equation}
where $\Lambda_\nu$ is the scale where the new physics generating this operator is decoupled. When the electroweak symmetry is broken, this leads to the following mass matrix:
\begin{equation}
\label{massmatrix} 
M^\nu_{ij}=-\frac{1}{4}\frac{m^\nu_{ij}}{\Lambda_\nu}\,v^2 ~,
\end{equation}
where $v$ is the vacuum expectation value of the Higgs field $H$.

In SUSY models, we can build a SUSY operator analogous to Eq.~(\ref{dim5}) by replacing all fermion fields by superfields, and the $H$ field by the SUSY $H_u$ superfield. This modifies the mass matrix of Eq.~(\ref{massmatrix}) by a factor $\sin^2\beta$. Note that this relies on the assumption that either $\Lambda_\nu$ is larger than the SUSY breaking scale, such that the corresponding dimension-five operator must be holomorphic, or that the non-holomorphic SUSY-breaking corrections are small.

Although at high scales the neutrino mass matrix is highly model-dependent (i.e., see-saw mechanisms~\cite{Minkowski:1977sc,Mohapatra:1979ia,Schechter:1980gr,Foot:1988aq}, radiative mass generation~\cite{Zee:1980ai,Babu:1988ki}, SUSY with R-Parity breaking~\cite{Hall:1983id}), at lower scales the situation is different. After assuming a value for the lightest neutrino mass and the scale $\Lambda_\nu$, one can uniquely define the dimension-five operator using only the neutrino masses and mixing parameters. This implies that, to study the RGE evolution of the neutrino parameters within models like the SM and MSSM, one can introduce the neutrino mass matrix operator at the electroweak scale and not worry about its origin until the moment one reaches the scale $\Lambda_\nu$.

Furthermore, this operator is the only dimension-five operator allowed by the gauge symmetries. Given the dimensionality and flavour structure of this operator, its renormalization must always be proportional to the operator itself, leading to the following RGE $\beta_\nu$ function~\cite{Antusch:2003kp}:
\begin{equation}
 16\pi^2\frac{d m^\nu_{ij}}{dx}=C\left((Y_e^\dagger Y_e)^T_{ik}m^\nu_{kj}+m^\nu_{ik}(Y_e^\dagger Y_e)_{kj}\right)+\alpha\,m^\nu_{ij} ~,
\end{equation}
where $x\equiv\ln(\mu/\mu_0)$, $\mu$ is the scale where the operator is being evaluated, and $\mu_0$ is the scale where the initial conditions are specified. Here, and in the following equations, $Y_f$ shall denote the Yukawa matrix of the fermion $f=u,d,e$. The coefficients $C$ and $\alpha$ were calculated in~\cite{Babu:1993qv,Antusch:2001ck,Antusch:2001vn}:
\begin{eqnarray}
 C & = & \left\{\begin{array}{lcr}
-\frac{3}{2} & & \textrm{(SM)} \\
\;\;\;1 & & \textrm{(MSSM)} 
\end{array}\right. \\
\alpha & = & \left\{\begin{array}{lcr}
-3g_2^2+2\,\textrm{Tr}(Y_e^\dagger Y_e)+6\,\textrm{Tr}(Y_u^\dagger Y_u)+6\,\textrm{Tr}(Y_d^\dagger Y_d)+\lambda & & \textrm{(SM)} \\
-\frac{6}{5}g_1^2-6g_2^2+6\,\textrm{Tr}(Y_u^\dagger Y_u) & & \textrm{(MSSM)}
\end{array}\right.
\end{eqnarray}

A study of the variation of the neutrino mixing parameters in the SM and MSSM was carried out in~\cite{Antusch:2003kp,Antusch:2005gp}. The authors found that, below the $\Lambda_\nu$ scale, the variation of the mixing angles could be approximated using the following analytical equations:
\begin{subequations}
\label{angdots}
\begin{eqnarray}
\label{dott12}
 \dot{\theta}_{12} & = & -C\frac{y_\tau^2}{32\pi^2}\sin2\theta_{12}\sin^2\theta_{23} \frac{\left|m_1\,e^{i\phi_1}+m_2\,e^{i\phi_2}\right|^2}{\Delta m^2_{21}} + \mathcal{O}(\theta_{13}) \\
\label{dott23}
 \dot{\theta}_{23} & = & -C\frac{y_\tau^2}{32\pi^2}\sin2\theta_{23}
\left(\cos^2\theta_{12}\frac{\left|m_2\,e^{i\phi_2}+m_3\right|^2}{\Delta m^2_{32}}+ \sin^2\theta_{12}\frac{\left|m_1\,e^{i\phi_1}+m_3\right|^2}{\Delta m^2_{31}}\right) + \mathcal{O}(\theta_{13}) \\
\label{dott13}
\dot{\theta}_{13} & = & C\frac{y_\tau^2}{32\pi^2}\sin2\theta_{12}\sin2\theta_{23}
\left(\frac{m_3m_1}{\Delta m^2_{31}}\cos(\phi_1-\delta)-\frac{m_3m_2}{\Delta m^2_{32}}\cos(\phi_2-\delta)- \frac{\Delta m^2_{21}}{\Delta m^2_{31}\Delta m^2_{32}}m_3^2\cos\delta\right) \nonumber \\
& & + \mathcal{O}(\theta_{13}) 
\end{eqnarray}
\end{subequations}
where $y_\tau=(Y_e)_{33}$ and $m_i$ are the neutrino masses. The full set of analytical equations can be found in~\cite{Antusch:2003kp}. Moreover, although these equations are useful for understanding the variations obtained by the running, in all our results we shall solve the exact equations using the \texttt{REAP} package~\cite{Antusch:2005gp}.

As these three equations are mainly dominated by $y_\tau^2$, in the MSSM we can expect an enhancement of order $(v/v_d)^2=(1+\tan^2\beta)$ with respect to the SM. Thus, for values of $\tan\beta\sim50$, we can get very large deviations in the mixing parameters at high scales. Moreover, in all models, the derivatives of the mixing angles are proportional to the neutrino masses, so the variations can be further enhanced if the neutrino masses are large. A further source of enhancement comes from the $\phi_i$ Majorana phases, which can conspire in order to increase the effects of the running even further.

As we want to investigate the largest possible variation of the neutrino flavour-transition probability with respect to its value evaluated at the best fit values, we shall maximize the $m_\nu$ and $\tan\beta$ enhancements. In this work we shall use a normal mass hierarchy, and set the lightest neutrino mass at $m_1=0.43$~eV, which is the upper limit allowed by the WMAP-only 7-year data~\cite{Komatsu:2010fb}. In addition, in the MSSM, we shall also use $\tan\beta=50$, and a SUSY decoupling scale $\Lambda_\text{SUSY}=1$~TeV.

\section{Neutrino oscillation probability and RGE effects}
\label{sec:setup}

The analytical expressions shown in eq.~(\ref{angdots}) are of considerable use for model builders, as they can connect low-energy observables in the neutrino sector with high-scale theories like Leptogenesis~\cite{Giudice:2003jh,Davidson:2008pf}, Flavour Symmetries~\cite{Varzielas:2008jm,Hirsch:2003dr,Lin:2009sq}, and Grand Unified Theories~\cite{Masiero:2005ua,Hisano:1998fj}. However, it is not clear if the effects of the running could be observed in an experimental situation or not.

As previously mentioned, when introducing quantum corrections to interaction vertices, the scale $\mu$ is set equal to the transferred momentum $Q=\sqrt{-q^2}$. On the other hand, deviations from the no-running scenario occur at scales larger than the SUSY decoupling scale. Thus, we need large values of $Q$. These can be obtained, among other conditions, when we have attained large values of the neutrino energy, $E_\nu$. In particular, if we assume $\Lambda_\text{SUSY}=1$~TeV, to get a $Q$ of this order of magnitude we shall require energies much larger than those found in present and future long-baseline neutrino experiments. For instance, the largest energy considered for neutrino factories is of about 50~GeV. Thus, in order to reach larger energies, we must use astrophysical neutrinos. Examples of sources of these types of neutrinos are supernova remnants~\cite{Costantini:2004ap}, gamma ray bursts~\cite{Waxman:1997ti}, and active galactic nuclei (AGN)~\cite{Stecker:1991vm}, with expected neutrino energies of order $10^{13}$, $10^{14}$ and $10^{18}$~eV, respectively. For definiteness, we restrict our analysis to AGN, which provide the neutrinos with the highest possible energy in the Universe.

In a neutrino experiment, there are two different moments where the scale is relevant: in the production and the detection of the neutrinos. In AGN, the neutrino production is realised though the decay of a particle, typically the pion, which sets a low scale for the transferred momentum (i.e., $Q=m_\pi$). Meanwhile, when the ultra high energy neutrinos reaching the Earth interact through deep inelastic scattering with the nucleons of the detector, there is a possibility to achieve a large value of transferred momentum. It is in this last context where we can meet the necessary conditions to test the RGE effects in the neutrino system. Therefore, and given that we have two different $Q$ scales, the $U_\nu$ matrix used to calculate the incoming neutrino flux will differ from the $U^\prime_\nu$ matrix that appears in the detection cross-section. We parametrise the $U^\prime_\nu$ matrix, as in eq.~(\ref{EqUPMNS}), by three $\zeta_{ij}$ mixing angles and a phase $\delta_1$, which are calculated by solving the RGE equations for the $\theta_{ij}$ mixing angles measured in the oscillation experiments. Due to the low scale involved in the neutrino production, the $U_\nu$ matrix is given by the current measurements of the neutrino oscillation parameters. These are shown in eq.~(\ref{mixing_angles}), and from now on shall be those used in our analysis.

As the $U_\nu$ and $U^\prime_\nu$ matrices can be factorised out of the production decay rate and detection cross section, we will be able to define a function $P_{\alpha\beta}$, which will measure the probability of a neutrino produced with flavour $\alpha$ interacting with flavour $\beta$:
\begin{equation}\label{EqProbPabQ}
 P_{\alpha\beta}(Q) = \sum_{i=1}^3{\Bigl|(U_\nu)_{\alpha i}\Bigr|^2\left|\left(U^\prime_\nu(Q)\right)_{\beta i}\right|^2} ~.
\end{equation}
Notice that if we set $U^\prime_\nu\to U_\nu$, this expression coincides with the standard expression for neutrino oscillation probability at large distances.

It is useful to have an unraveled expression for this new probability. In the following, for conciseness, we shall write $g_{\alpha i}=\left|(U_\nu)_{\alpha i}\right|^2$. Making the dependence on the $\zeta_{ij}$ angles and the CP phase $\delta_1$ explicit, we can write the probabilities as:
\begin{eqnarray}
 \label{EqPbetae}
 P_{\alpha e} 
 &=& 
 g_{\alpha 3} + c_{\zeta_{13}}^2 \left[ \left(g_{\alpha 1}-g_{\alpha 3}\right) - s_{\zeta_{12}}^2 \left(g_{\alpha 1}-g_{\alpha 2}\right) \right] ~, \\
 \label{EqPbetamu}
 P_{\alpha\mu}
 &=&
 g_{\alpha 2} 
 + s_{\zeta_{12}}^2 \left(g_{\alpha 1}-g_{\alpha 2}\right) \nonumber \\
 && + ~s_{\zeta_{23}}^2 \left[ \left(g_{\alpha 3}-g_{\alpha 2}\right) - s_{\zeta_{12}}^2 \left(1+s_{\zeta_{13}}^2\right) \left(g_{\alpha 1}-g_{\alpha 2}\right) + s_{\zeta_{13}}^2 \left(g_{\alpha 1}-g_{\alpha 3}\right) \right] \nonumber \\
 && + \left(\frac{g_{\alpha 1}-g_{\alpha 2}}{2}\right) 
 c_{\delta_1} s_{2\zeta_{12}} s_{2\zeta_{23}} s_{\zeta_{13}} ~,\\
 \label{EqPbetatau}
 P_{\alpha\tau}
 &=&
 g_{\alpha 2} 
 + s_{\zeta_{12}}^2 \left(g_{\alpha 1}-g_{\alpha 2}\right) \nonumber \\
 && + ~c_{\zeta_{23}}^2 \left[ \left(g_{\alpha 3}-g_{\alpha 2}\right) - s_{\zeta_{12}}^2 \left(1+s_{\zeta_{13}}^2\right) \left(g_{\alpha 1}-g_{\alpha 2}\right) + s_{\zeta_{13}}^2 \left(g_{\alpha 1}-g_{\alpha 3}\right) \right] \nonumber \\
 && - \left(\frac{g_{\alpha 1}-g_{\alpha 2}}{2}\right) 
 c_{\delta_1} s_{2\zeta_{12}} s_{2\zeta_{23}} s_{\zeta_{13}} ~,
\end{eqnarray}
with $c_\chi \equiv \cos \chi$ and $s_\chi \equiv \sin \chi$. The explicit dependence of the $g_{\alpha i}$ on the standard mixing angles can be found in Appendix \ref{AppendixA}. Hereafter, we will consider only oscillation probabilities of the form $P_{e\beta}$ and $P_{\mu\beta}$ ($\beta = e, \mu, \tau$), since the electron and muon neutrinos are the only ones that intervene in the different models of the neutrino production at AGN.  

Naturally, the probabilities do not have an explicit dependence on the Majorana phases. However, as the final value of the $\zeta_{ij}$ angles depends on the value of each phase, we find that there does exist an implicit dependence on them. In particular, we have found that the higher the value of the Majorana phase difference $\phi_1-\phi_2$, the higher the possible values of $\zeta_{12}$ and $\zeta_{13}$ that are respectively achievable. In $\zeta_{23}$ there is a different behaviour: as $\phi_1-\phi_2$ rises, the accessible region of $\zeta_{23}$ becomes focalised around a certain value, dependent on the chosen values for the mixing parameters at production. Note that, as long as $\delta=0$, the regions are the same for $\phi_1-\phi_2$ and for $\phi_2-\phi_1$. Nevertheless, using $\delta \neq 0$ does not generate large deviations, and these statements remain valid.

\begin{figure}[t!]
 \begin{center}
  \scalebox{0.5}{\includegraphics{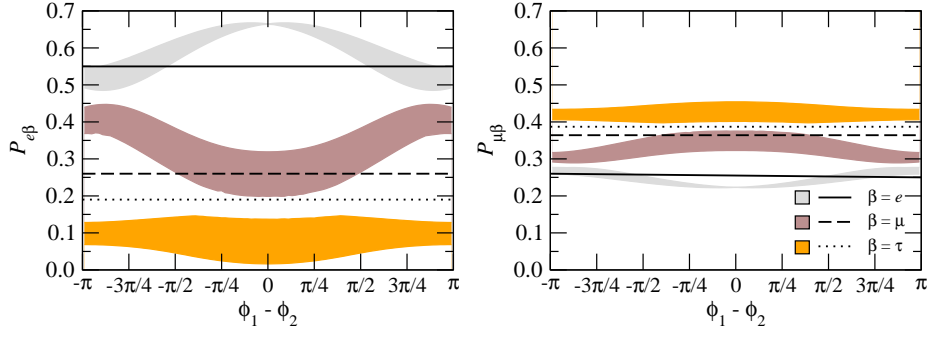}} 
  \caption{\label{FigProbVsPhiDiff} Flavour-transition probabilities $P_{e\beta}$, $P_{\mu\beta}$ versus difference between the Majorana phases, $\phi_1-\phi_2$. Lines correspond to standard probabilities, with $\beta = e$ (solid), $\mu$ (dashed), and $\tau$ (dotted). Areas correspond to modified probabilities, where, at each point, $\phi_1$ has been randomly varied, while $\phi_2$ has been adjusted to keep the desired phase difference. We have $\beta = e$ (grey), $\mu$ (brown), and $\tau$ (orange). The RGE evolution was performed at $Q^2 = 10^{11}$ GeV$^2$, fixing the $\theta_{ij}$ and $\Delta m_{ij}^2$ at their best-fit values, and $\delta = 0$.}
 \end{center}
\end{figure}

We find it illustrative to show how the probabilities depend on the Majorana phase difference, which is shown in figure \ref{FigProbVsPhiDiff}. Here, the left and right panels show, respectively, the probabilities $P_{e\beta}$ and $P_{\mu\beta}$, for $\beta = e$ (solid line, grey areas), $\mu$ (dashed line, brown areas), and $\tau$ (dotted line, orange areas). The lines correspond to the standard, no-running, probabilities, while the areas correspond to the probabilities influenced by the MSSM running. To obtain the areas, we have fixed all parameters at their best-fit values, and set $\delta=0$, $Q^2 = 10^{11}$ GeV$^2$. For each phase difference, we have varied $\phi_1$ between $0$ and $2\pi$, and adjusted $\phi_2$ to keep the difference at the desired value.

Figure \ref{FigProbVsPhiDiff} shows a strong dependence of the transition probability on the Majorana phases, in clear distinction with the standard neutrino oscillation scenario. In fact, after the masses, the Majorana phases are the most important parameters in the neutrino mass operator affecting the final value of the probability. Note that the maximum deviation of the modified $P_{ee}$ and $P_{e\tau}$ occurs at $\phi_1-\phi_2 = 0$, while the maximum deviation of $P_{e\mu}$ occurs at $\phi_1-\phi_2 = \pm\pi$. The deviations are less pronounced for the $P_{\mu\beta}$ probabilities, with $P_{\mu e}$ and $P_{\mu\tau}$ reaching maximum deviation at $\phi_1-\phi_2=0$, and $P_{\mu\mu}$ at $\phi_1-\phi_2=\pm\pi$.

\begin{figure}[t!]
  \begin{center}
    \scalebox{0.55}{\includegraphics{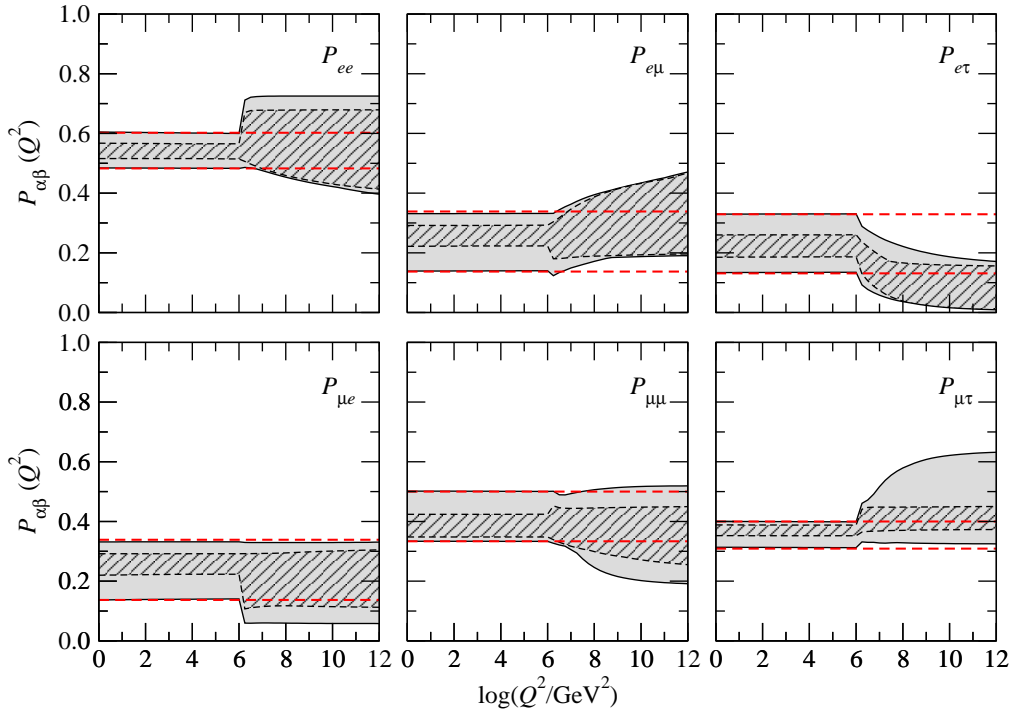}} 
    \caption{Flavour-transition probabilities under MSSM running, as functions of the transferred momentum $Q^2$. In the gray areas we vary all mixing parameters at production within their current $3\sigma$ bounds (the CP and Majorana phases were varied between $0$ and $2\pi$), while in the hatched areas only $\theta_{13}$ and the phases are varied. The dashed (red) bands indicate the limits corresponding to standard oscillations.}
    \label{FigProbvsQ2}
  \end{center}
\end{figure}

It is also instructive to understand how the transition probabilities depend on the scale $\mu=Q$, which is shown in figure \ref{FigProbvsQ2}. In this plot, the gray areas represent the accessible values of the probability after MSSM running when the standard mixing angles and squared-mass differences are varied within their current $3\sigma$ experimental bounds, and the phases are varied between $0$ and $2\pi$. The hatched regions show the same, but considering only the variation of $\theta_{13}$ and the phases. From this we can understand the role played by the uncertainty in the other better-known parameters. The bands limited by dashed (red) lines correspond to the probability calculated using only standard oscillations, without MSSM running, when the mixing parameters are allowed to vary in the same manner. Notice that the deviations in the probabilities start at $Q^2 = 10^6$~GeV$^2 =\Lambda^2_\text{SUSY}$ and increase with $Q^2$.

Figure \ref{FigProbvsQ2} allows us to make definite predictions regarding each probability. Although the high-scale behaviour of $P_{ee}$ is uncertain, we can expect $P_{e\mu}$ to generally increase, and $P_{e\tau}$ to always decrease with $Q^2$. In fact, we can obtain a suppression of $P_{e\tau}$ down to zero, which shall provide very interesting results in the next sections. This behaviour is consistent with a large decrease of $\zeta_{23}$, with deviations from the general trend depending on the value of the Majorana phases, as shown in figure \ref{FigProbVsPhiDiff}.

On the other hand, $P_{\mu\tau}$ tends to increase with $Q^2$. $P_{\mu e}$ and $P_{\mu\mu}$ shall generally be expected to decrease, although it is possible for them to remain invariant. This again is consistent with a decrement of $\zeta_{23}$, and deviations depending on the exact value of the Majorana phases.

\section{Astrophysical observables}
\label{sec:theory}

\subsection{The UHE astrophysical neutrino flux}\label{Section_Theory_Sub_UHEflux}

AGN have long been presumed to be sites of ultra-high-energy (UHE) neutrino production. In the scenario of neutrino production by meson decay, it is assumed that within the AGN protons are accelerated through first-order Fermi shock acceleration~\cite{Kachelriess:2008ze} and that pions are produced in the processes
\begin{equation}\label{EqPionProd}
 p + \gamma \rightarrow \Delta^+ \rightarrow
 \left\{\begin{array}{l}
 p + \pi^0 \\
 n + \pi^+
 \end{array}\right. \quad, \qquad
 n + \gamma \rightarrow p + \pi^- ~.
\end{equation}
The neutral pions decay into gamma rays through $\pi^0 \rightarrow \gamma \gamma$, while the charged pions decay into electron- and muon-neutrinos through
\begin{equation}\label{EqPionDecay}
 \pi^+ \rightarrow \nu_\mu + \mu^+ \rightarrow \nu_\mu + e^+ + \nu_e + \overline{\nu}_\mu \quad, \qquad
 \pi^- \rightarrow \overline{\nu}_\mu + \mu^- \rightarrow \overline{\nu}_\mu + e^- + \overline{\nu}_e + \nu_\mu ~.
\end{equation}
If neutrinos are produced by these processes, then the ratios of the different flavours ($\nu_x + \overline{\nu}_x$) to the total flux are:
\begin{equation}\label{EqRatiosProd}
 \left(\Phi_{\nu_e+\bar{\nu}_e}^0 : \Phi_{\nu_\mu+\bar{\nu}_\mu}^0 : \Phi_{\nu_\tau+\bar{\nu}_\tau}^0\right) = \left(1/3 : 2/3 : 0\right) ~.
\end{equation}
Note, however, that these flavour fluxes are approximate; a detailed analysis of the pion decay and the contribution from higher-energy processes will nevertheless result in values that are close to these standard ratios \cite{Lipari:2007su,Athar:2005wg}.

By the time neutrinos reach Earth, neutrino oscillations will have redistributed the flux among the three flavours, so that the flavour fluxes at detection are
\begin{equation}\label{EqDetFlavRatios}
 \Phi_{\overset{\brabar}{\nu}_\alpha} = \sum_{\beta=e,\mu,\tau} P_{\beta\alpha} \Phi_{\overset{\brabar}{\nu}_\beta}^0 ~.
\end{equation}
For instance, standard mass-driven neutrino oscillations, evaluated at the best-fit values of the mixing parameters, eq.~(\ref{mixing_angles}), distribute the total flux in the $\pi^\pm$ decay scenario in an approximately uniform manner among the three flavours, i.e., $\left(\Phi_{\nu_e+\bar{\nu}_e} : \Phi_{\nu_\mu+\bar{\nu}_\mu} : \Phi_{\nu_\tau+\bar{\nu}_\tau}\right) \approx \left(0.36 : 0.33 : 0.31\right)$.

Deviations from these values could signal the presence of new physics at work. In fact, there are several theoretical hypotheses, such as violation of Lorentz or CPT invariance~\cite{Barenboim:2003jm,Bustamante:2010nq}, that could in principle work on top of the standard mass-driven oscillation mechanism and induce these deviations. In particular, as we have shown before, the MSSM introduces RGE-induced changes in the standard neutrino oscillation probability, which implies that such deviations could be also achieved within this scenario. However, the expected flavour fluxes could be also modified by using a different neutrino production mechanism. Thus, it is essential to thoroughly describe the deviations we would expect from the MSSM, including simultaneously the different potential production processes we know of. Note that, while it is possible for the flavour ratios at production to have an energy dependence \cite{Hummer:2010ai}, we have not considered this possibility in our analysis.

In a related production process \cite{Lipari:2007su,Rachen:1998fd,Kashti:2005qa,Hummer:2010ai}, the muons produced by pion decay may lose most of their energy before decaying, so that a pure-$\overset{\brabar}{\nu}_\mu$ flux is generated at the source, i.e.,
\begin{equation}
 \left(\Phi_{\nu_e+\bar{\nu}_e}^0:\Phi_{\nu_\mu+\bar{\nu}_\mu}^0:\Phi_{\nu_\tau+\bar{\nu}_\tau}^0\right) = \left(0:1:0\right) ~.
\end{equation}
Under standard oscillations, and with the mixing parameters set at their best-fit values, these initial fluxes yield, at Earth, $\left(\Phi_{\nu_e+\bar{\nu}_e}:\Phi_{\nu_\mu+\bar{\nu}_\mu}:\Phi_{\nu_\tau+\bar{\nu}_\tau}\right) \approx \left(0.26:0.36:0.38\right)$. 

Alternatively, a pure-$\bar{\nu}_e$ initial flux, i.e.,
\begin{equation}
 \left(\Phi_{\bar{\nu}_e}^0:\Phi_{\nu_\mu+\bar{\nu}_\mu}^0:\Phi_{\nu_\tau+\bar{\nu}_\tau}^0\right) = \left(1:0:0\right) ~,
\end{equation}
has been considered, e.g., in \cite{Lipari:2007su,Hummer:2010ai}. In this scenario, high-energy nuclei emitted by the source have sufficient energy for photodisintegration to occur, but not enough to reach the threshold for pion photoproduction. The neutrons created in the process generate $\overline{\nu}_e$ through beta decay. Using standard oscillations and best-fit values for the mixing parameters, this yields, at Earth, $\left(\Phi_{\bar{\nu}_e}:\Phi_{\bar{\nu}_\mu}:\Phi_{\bar{\nu}_\tau}\right) \approx \left(0.55:0.26:0.19\right)$.

Finally, semileptonic decays of charm quarks can generate flavour fluxes at production of \cite{Enberg:2008te,Choubey:2009jq}
\begin{equation}
 \left(\Phi_{\nu_e+\bar{\nu}_e}^0:\Phi_{\nu_\mu+\bar{\nu}_\mu}^0:\Phi_{\nu_\tau+\bar{\nu}_\tau}^0\right) = \left(1/2:1/2:0\right) ~.
\end{equation}
These fluxes can also be produced as the result of a pile-up effect \cite{Hummer:2010ai}. In this scenario, standard oscillations at best-fit values result in $\left(\Phi_{\nu_e+\bar{\nu}_e}:\Phi_{\nu_\mu+\bar{\nu}_\mu}:\Phi_{\nu_\tau+\bar{\nu}_\tau}\right) \approx \left(0.41:0.31:0.28\right)$.  

The initial neutrino and anti-neutrino flavour fluxes for the four production scenarios that we have considered are summarised in table \ref{TblScenarios}. The fluxes at Earth for the four scenarios are hence:
\begin{eqnarray}
 \label{EqFlavRatiosScA}
 \text{Scenario A}&:&
 \left\{\begin{array}{l}
  \Phi_{\nu_\alpha}\left(E_\nu\right) = 0 \\
  \Phi_{\bar{\nu}_\alpha}\left(E_\nu\right) = P_{e\alpha}\left(E_\nu\right) \\
 \end{array}\right. \\
 \label{EqFlavRatiosScB}
 \text{Scenario B}&:& 
 \Phi_{\overset{\brabar}{\nu}_\alpha}\left(E_\nu\right)
  = \frac{1}{2} P_{\mu\alpha}\left(E_\nu\right) \\
 \label{EqFlavRatiosScC}
 \text{Scenario C}&:& 
 \Phi_{\overset{\brabar}{\nu}_\alpha}\left(E_\nu\right)
  = \frac{1}{4} \left( P_{e\alpha}\left(E_\nu\right)
                       + P_{\mu\alpha}\left(E_\nu\right) \right) \\
 \label{EqFlavRatiosScD}
 \text{Scenario D}&:& 
 \Phi_{\overset{\brabar}{\nu}_\alpha}\left(E_\nu\right)
  = \frac{1}{6} \left( P_{e\alpha}\left(E_\nu\right)
                       + 2 P_{\mu\alpha}\left(E_\nu\right) \right)
\end{eqnarray}

\begin{table}[t!]
 \begin{center}
   \begin{tabular}{|c|l|c|c|}
    \hline
    Scenario & Description & $\left(\Phi_{\nu_e}^0:\Phi_{\nu_\mu}^0:\Phi_{\nu_\tau}^0\right)$ 
       & $\left(\Phi_{\bar{\nu}_e}^0:\Phi_{\bar{\nu}_\mu}^0:\Phi_{\bar{\nu}_\tau}^0\right)$ \\
    \hline
    A & Beta decay   & $\left(0:0:0\right)$     & $\left(1:0:0\right)$ \\
    B & Muon-damped  & $\left(0:1/2:0\right)$   & $\left(0:1/2:0\right)$ \\
    C & Charm decay  & $\left(1/4:1/4:0\right)$ & $\left(1/4:1/4:0\right)$ \\
    D & Pion decay   & $\left(1/6:1/3:0\right)$ & $\left(1/6:1/3:0\right)$ \\
    \hline
   \end{tabular}
 \end{center}
 \caption{\label{TblScenarios} The four scenarios of neutrino flavour ratios at production time. The normalised fluxes $\Phi_{\nu_\alpha}^0$ and $\Phi_{\bar{\nu}_\alpha}^0$ represent, respectively, the fraction of neutrinos and anti-neutrinos of flavour $\alpha$ produced at the source. Note that, except for scenario A, the flux is equally divided between particles and anti-particles.}
\end{table}

\subsection{Flavour ratios}\label{SecTheorySubFlavRatios}

Although we have four well-motivated models giving the flavour composition of the initial neutrino flux coming from AGN, the total neutrino flux is subject to many uncertainties (see~\cite{Anchordoqui:2005is} and references within). A good way to avoid having to deal with the uncertainties in the initial neutrino flux is using ratios of fluxes measured on Earth~\cite{Beacom:2003nh}. For instance, a typical ratio used is
\begin{equation}\label{EqDefT}
 T
 = \frac{\Phi_{\nu_\mu+\bar{\nu}_\mu}}{\Phi_{\nu_e+\bar{\nu}_e}+\Phi_{\nu_\mu+\bar{\nu}_\mu}+\Phi_{\nu_\tau+\bar{\nu}_\tau}} 
 = \Phi_{\nu_\mu+\bar{\nu}_\mu} ~,
\end{equation}
which is related to the experimental ratio of $\nu_\mu$ \u{C}erenkov events to the total number of neutrino events. If it is possible to distinguish between $\nu_e$ and $\nu_\tau$ showers, another convenient ratio is
\begin{equation}\label{EqDefR}
 R=\frac{\Phi_{\nu_e+\bar{\nu}_e}}{\Phi_{\nu_\tau+\bar{\nu}_\tau}} ~.
\end{equation}

In what follows, we will show our expectations for $R$ and $T$ considering the four different scenarios, A, B, C and D, and assuming no running and MSSM running of the mixing parameters.

In all these results, we variate $\theta_{13}$ in its $3\sigma$ range, and all phases between $0$ and $2\pi$. We will either hold all other oscillation parameters at their best-fit values, or variate them in their $3\sigma$ range. Also, throughout the present section we shall fix $Q^2=10^{11}$~GeV$^2$ in order to maximise the MSSM RGE effects.

\subsubsection{Scenario A: Production through beta decay}

\begin{figure}[t]
\centering
\includegraphics[scale=.6]{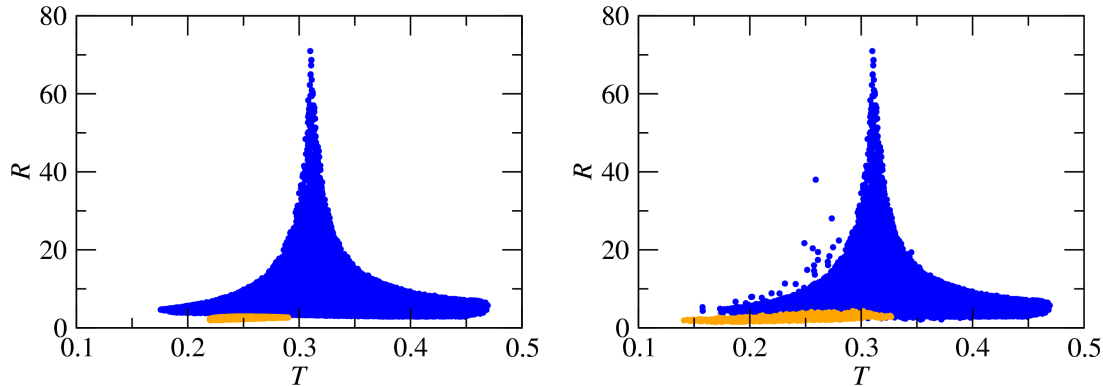}
\caption{Expected ratio of fluxes assuming no running (orange) and MSSM RGE running (blue) of mixing parameters at $Q^2 = 10^{11}$ GeV$^2$, for a $(1:0:0)$ production model. We show only the variation of $\theta_{13}$ and the phases on the left, and the variation of all oscillation parameters on the right, as described in the text.}
\label{fig:100}
\end{figure}

For simplicity, we start with the initial $(1:0:0)$ ratio, which occurs if the neutrino production at AGN is governed by beta decay. We then have:
\begin{equation}
 T =  P_{e\mu} \quad, \quad
 R = \frac{P_{e e}}{P_{e\tau}} ~.
\end{equation}

From figure \ref{FigProbvsQ2}, we would generally expect an enhancement in $P_{e\mu}$, such that $T$ is larger than the no-running typical values. Also, although whether $P_{ee}$ is enhanced or suppressed depends on the values of the oscillation parameters at production time, we have seen that $P_{e\tau}$ is generally suppressed, with the possibility of being zero. Thus, we expect $R$ to be able to reach very large values.

In figure \ref{fig:100} we show the scatter plot of $R$ vs.~$T$. There we observe that the no-running expectations at $3\sigma$ are $T\in(0.14,0.33)$ and $R\in(1,4)$, while the MSSM running expands this into $T\in(0.15,0.47)$ and $R\in(2,70)$. We see that it is possible for $R$ to reach up to 18 times the highest possible no-running value, a clear indication of new physics. $T$ can also become very large, being particularly useful if the errors in other oscillation parameters are diminished. Note that the ranges of $R$ and $T$ can be extracted in a straightforward manner due to the directly or inversely proportional relationship between these observables and the oscillation probabilities. For instance, the upper (lower) limit of $R$ has been obtained from the ratio of the maximum and minimum values of the oscillation probabilities in figure ~\ref{FigProbvsQ2}, $P_{ee}/P_{e\mu} \approx 0.72/0.01 ~\left(0.4/0.2\right)$, at $Q^2=10^{11}$~GeV$^2$.

It is interesting to note that since the probabilities shown in figure~\ref{FigProbvsQ2} are moderately flat around $Q^2=10^{11}$~GeV$^2$, we can expect these results to hold even for somewhat lower values of $Q^2$. This particular observation applies to the other three production scenarios as well.

It can also be seen that the correlation of $T$ and $R$ is invaluable. Apart from being indicators of new physics, both ratios define a zone in which the new physics effect can be identified as coming from SUSY. If either $R$ or $T$ are observed out of both orange and blue regions, the new physics origin of the deviations could not be attributed to MSSM RGEs.

\subsubsection{Scenario B: Muon-damped production}

\begin{figure}[t]
\centering
\includegraphics[scale=.6]{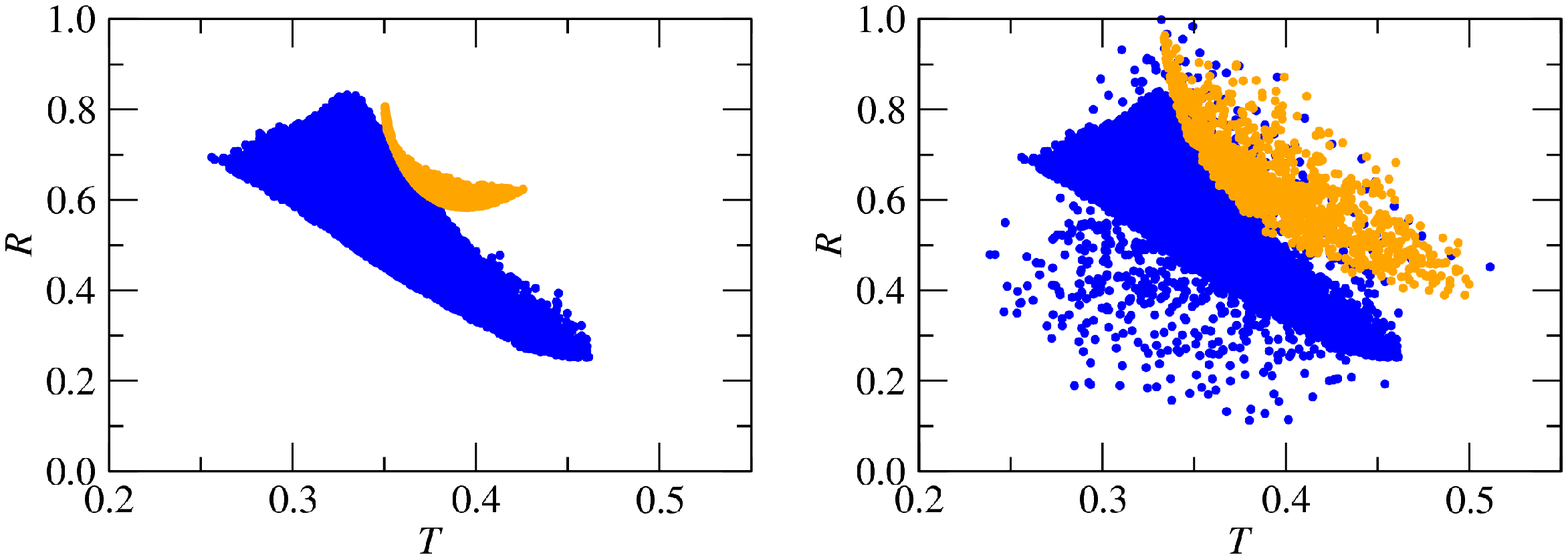} 
\caption{Same as figure \ref{fig:100}, but for a $(0:1:0)$ production model.}
\label{fig:010}
\end{figure}

The muon-damped production process generates the $(0:1:0)$ fluxes. The $T$ and $R$ ratios can then be written as:
\begin{equation}
 T = P_{\mu\mu} \quad, \quad
 R = \frac{P_{\mu e}}{P_{\mu\tau}} ~.
\end{equation}

We can again use figure \ref{FigProbvsQ2} to predict how these ratios shall behave. In contrast to the previous production process, we could expect some suppression in $P_{\mu\mu}$, giving a value of $T$ lower than the typical ones. Furthermore, the possible suppression of $P_{\mu e}$ and the enhancement of $P_{\mu\tau}$ would also decrease $R$, giving a situation opposite to the $(1:0:0)$ ratio.

Figure \ref{fig:010} illustrates the situation. The no-running expectations at $3\sigma$ are $T\in(0.33,0.5)$ and $R\in(0.4,1.0)$, while the MSSM running takes this into $T\in(0.25,0.5)$ and $R\in(0.1,1.0)$. As in the previous case, these ranges can be read directly from the corresponding range of $P_{\mu\mu}$, for $T$, and the proper combination of the maximum and minimum values of $P_{\mu\mu}$ and $P_{\mu\tau}$, for $R$.

In this case, the suppression of $T$ and $R$ is not as strong as their enhancement in the previous scenario. However, when correlated, it is clear that they can be disentangled from the no-running regions. Furthermore, it is also clear that the zone delimited by their correlation can be separated from the zone delimited in Scenario A. It should then be possible to deduce the original neutrino production process at the AGN.

\subsubsection{Scenario C: Production through charm decay}

\begin{figure}[t]
\centering
\includegraphics[scale=.6]{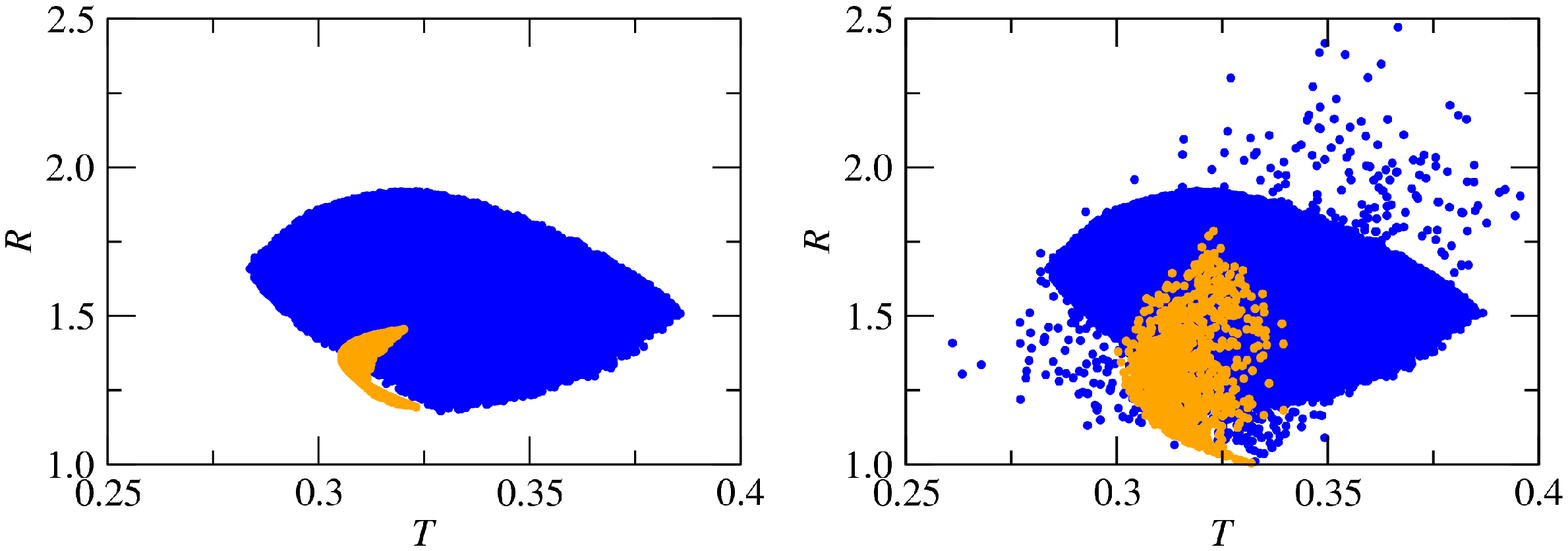}
\caption{Same as figure \ref{fig:100}, but for a $(1/2:1/2:0)$ production model.}
\label{fig:110}
\end{figure}

The decay of $D$ mesons can lead to a $(1/2:1/2:0)$ production ratio. The detected ratios are then described by:
\begin{equation}
 T = \frac{1}{2}(P_{e\mu}+P_{\mu\mu}) \quad, \quad
 R = \frac{P_{ee}+P_{\mu e}}{P_{e\tau}+P_{\mu\tau}} ~.
\end{equation}

The results for $T$ and $R$ can thus be understood as an average of the predictions for muon-damped and beta decay production. As we have seen, these two scenarios have opposite behaviours, so it is possible for the deviations to cancel each other. In fact, for $T$, the possible reduction of $P_{\mu\mu}$ and increase of $P_{e\mu}$ cause the total prediction to smooth out, getting milder maximum variations. For $R$, although $P_{e\tau}$ is decreased, the large $P_{\mu\tau}$ does not allow $R$ to reach the high values found in the beta decay production model. Furthermore, the smaller $P_{\mu e}$ combines with the somewhat larger $P_{ee}$, such that the reduction characteristic of the muon-damped production model is not observed. All in all, $R$ does not differ much from the no-running prediction.

We show the no-running and MSSM predictions in figure \ref{fig:110}. The no-running expectations at $3\sigma$ are $T\in(0.30,0.34)$ and $R\in(1.0,1.8)$, while the MSSM running expands this into $T\in(0.26,0.39)$ and $R\in(1.0,2.5)$. We can see it is unlikely for neither $T$ nor $R$ to be used to distinguish new physics from no-running behaviour. By reducing the uncertainty in the mixing parameters one could hope to improve the situation, as the no-running region becomes very tiny, but, still, the difference between the no-running and MSSM running regions will not be as large as for scenario A.

\subsubsection{Scenario D: Production through pion decay}

\begin{figure}[t]
\centering
\includegraphics[scale=.6]{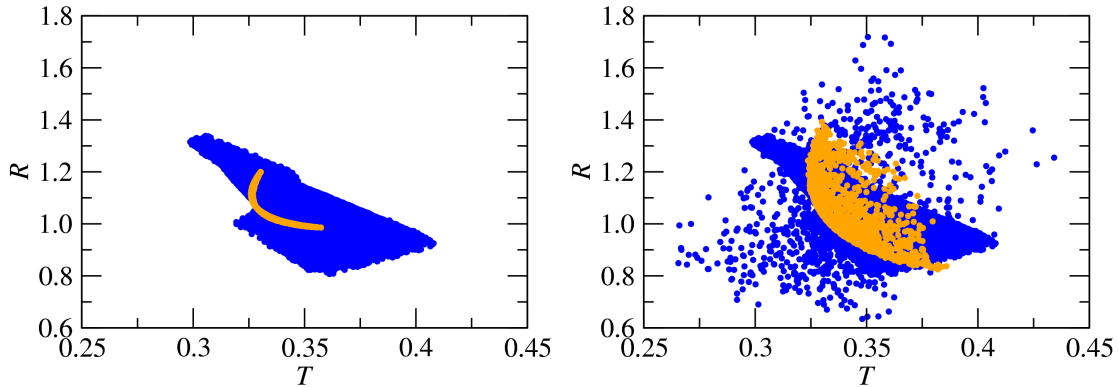}
\caption{Same as figure \ref{fig:100}, but for a $(1/3:2/3:0)$ production model.}
\label{fig:120}
\end{figure}

The $(1/3:2/3:0)$ model associated to full pion decay is arguably the one most studied in the literature. The ratios detected at Earth are described using:
\begin{equation}
 T = \frac{1}{3}(P_{e\mu}+2P_{\mu\mu}) \quad, \quad
 R = \frac{P_{ee}+2P_{\mu e}}{P_{e\tau}+2P_{\mu\tau}} ~.
\end{equation}

This is similar to the charm decay model, but since more weight is given to $P_{\mu\mu}$, $P_{\mu e}$ and $P_{\mu\tau}$, we can expect our results to lean slightly towards those for the muon-damped production model. Nonetheless, the variations are still not as significant as in the muon-damped and beta decay production models.

This is what we observe in figure \ref{fig:120}. The $3\sigma$ no-running expectations are $T\in(0.32,0.38)$ and $R\in(0.8,1.4)$, while the MSSM running expands this into $T\in(.27,.43)$ and $R\in(0.7,1.7)$. No large deviations from the no-running expectations can be observed in any parameter, even if there is an improvement in the measurement of oscillation parameters, unless a large resolution in both $T$ and $R$ is achievable experimentally.

\subsubsection{Other Production Scenarios}

\begin{figure}
\centering
\includegraphics[scale=.6]{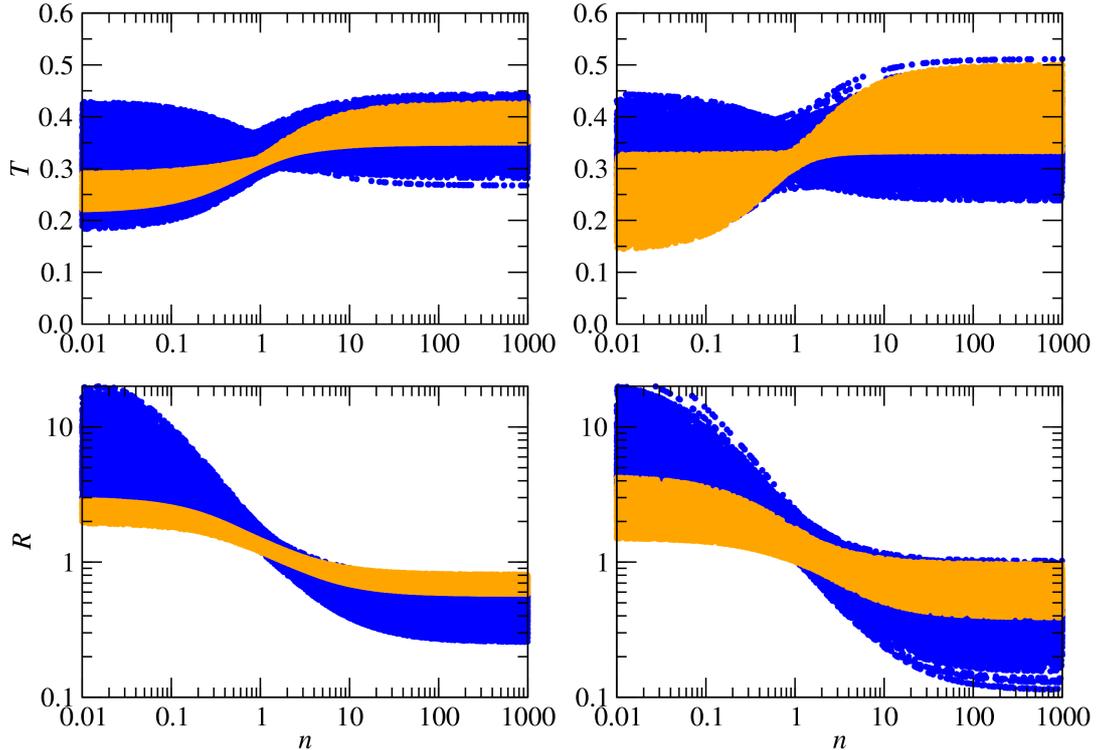}
\caption{$T$ (top) and $R$ (bottom) as a function of $n$ assuming SM (orange) and MSSM (blue) running of mixing parameters. We show the best fit values (left) and full $3\sigma$ variation (right), as described in the text.}
\label{fig:Tn}
\end{figure}

In the following we shall parametrise the initial $\left(\Phi^0_{\nu_e+\bar{\nu}_e}:\Phi^0_{\nu_\mu+\bar{\nu}_\mu}:\Phi^0_{\nu_\tau+\bar{\nu}_\tau}\right)$ flux as $(1:n:0)$, as in~\cite{Choubey:2009jq}. We show $T$ and $R$ as functions of $n$ in figure \ref{fig:Tn}, where we vary $n$ from $10^{-2}$ to $10^3$ in order to properly reproduce the limiting cases $(1:0:0)$ and $(0:1:0)$. Again, the orange region shows the no-running prediction for $T$ and $R$, while the blue region shows that for the MSSM RGEs.

From figure~\ref{fig:Tn} we see that the previously analysed scenarios are representative cases when varying $n$, since the $(1:2:0)$, $(0:1:0)$, $(1:0:0)$, and $(1:1:0)$ initial ratios correspond to $n=2$, $n=\infty$, $n=0$ and $n=1$, respectively.

However, figure~\ref{fig:Tn} allows us to understand additional features of our framework. For instance, we can see that the no-running scenario is bounded. In particular, taking into account the $3\sigma$ variation of all mixing parameters, $T$ is roughly bounded within $(0.15,0.5)$ and $R$ within $(0.4,4)$. Any measurement of these parameters outside these bounds is a clear indication of new physics.

Moreover, it is straightforward to see that the MSSM and no-running predictions for $T$ cannot be disentangled if $n$ is not known. A large deviation from the no-running scenario at $n\to0$ is degenerate to a situation with no deviation at $n\to\infty$, and vice-versa. Thus, on its own, $T$ cannot provide any useful information for our purpose.

In contrast, $R$ unambiguously shows very large deviations from the no-running scenario whenever the condition $|\log_{10}(n)|\gtrsim1$ is satisfied. In addition, the measurement of $R$ not only can reveal the presence of new physics in the running of neutrino mixing angles, but it can also indicate the initial composition of the neutrino flux coming from AGN. For instance, if we measure $R=10$, we not only know that there is new physics at work, but we can also set a limit of $n \lesssim 0.2$. Once information about $n$ is obtained, a correlation with $T$ is possible. In this case, the combination of both parameters can hint towards our MSSM RGE scenario.

\subsection{Average $R$ and $T$}

Deep inelastic neutrino-nucleon scattering (DIS), either by charged (CC) or neutral current (NC), is the most likely process to take place when an UHE astrophysical neutrino from AGN, with energy in the range $10^5 \leq E_\nu/\text{GeV} \leq 10^{12}$, interacts with the nucleons in the Earth. As a way to give a more realistic approach to our results we have calculated averaged forms of $R$ and $T$ using the DIS detection cross section.

In order to introduce the averaged forms of the flavour ratios, we have defined an averaged transition probability, weighed by the differential CC DIS cross section. This is given by the following expression:
\begin{equation}\label{EqProbAvg}
  \langle P_{\alpha\beta}\left(Q_\text{th}^2\right) \rangle^{\overset{\brabar}{\nu}}
 = \frac{1}{\sigma_\text{CC}^{\overset{\brabar}{\nu}}\left(E_\nu\right)} 
   \int_0^1 dx \int_0^1 dy \frac{d^2\sigma_\text{CC}^{\overset{\brabar}{\nu}}}{dxdy}\left(E_\nu,x,y\right) P_{\alpha\beta}\left(Q^2\right) ~,
\end{equation}
with the integrated CC cross section defined as
\begin{equation}\label{EqSigmaTot}
 \sigma_\text{CC}^{\overset{\brabar}{\nu}}\left(E_\nu\right)
 = \int_0^1 dx \int_0^1 dy \frac{d^2\sigma_\text{CC}^{\overset{\brabar}{\nu}}}{dxdy}\left(E_\nu,x,y\right) ~.
\end{equation}
We have used the standard formulae for the differential DIS cross sections \cite{Giunti:2007ry,Gandhi:1995tf}, with the CTEQ6M parton distribution functions\footnote{The pdf's are defined for $10^{-6} \leq x \leq 1$ and $\left(1.3\right)^2 \leq Q^2/\text{GeV}^2 \leq 10^8$; outside these ranges, they have been extrapolated \cite{CTEQ6web}.} (pdf's), fitted within the $\overline{MS}$ scheme \cite{Pumplin:2002vw}. The transferred momentum in a DIS event is $Q^2 = 2 m_N x y E_\nu$, with $x$ and $y$ the Bjorken scaling parameters\footnote{The Bjorken scaling parameters are defined as $x \equiv \frac{Q^2}{2 p_N \cdot q}$ and $y \equiv \frac{p_N \cdot q}{p_N \cdot p_\nu}$, with $p_N$ and $m_N$ the nucleon momentum and mass, respectively.}. Its value lies inside the range $0 \leq Q^2 \leq 2 m_N E_\nu$, the lower limit corresponding to $x=y=0$, and the upper limit, to $x=y=1$.

We find that, once we include the pdf's, the average probability favours values of low transferred momentum, erasing any effects due to the running. To avoid this, we introduce a cut-off in the form of a Heaviside function $\Theta\left(Q^2-Q_\text{th}^2\right)$ within the integrals in eq.~(\ref{EqProbAvg}). This artificially restricts the values of $Q^2$ to lie above $Q_\text{th}^2$, in an attempt to isolate the region of parameter space with high values of $Q^2$, where the MSSM-running effects are non-negligible.

\begin{figure}[t!]
  \begin{center}
    \scalebox{0.55}{\includegraphics{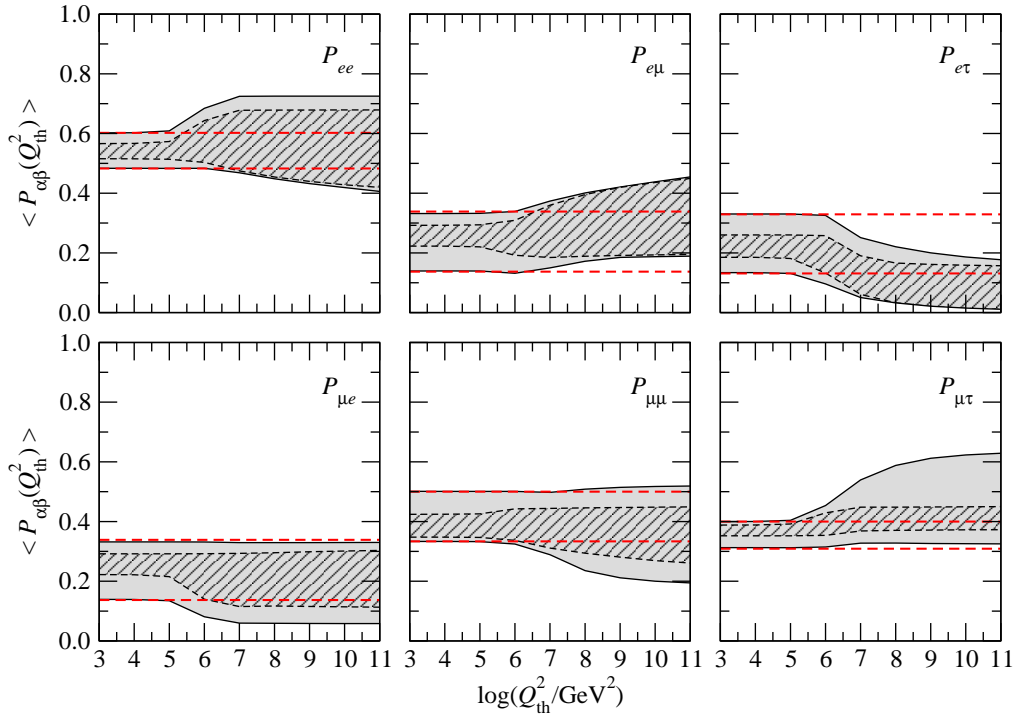}} 
    \caption{Cross-section-averaged flavour-transition probabilities under MSSM running, eq.~(\ref{EqProbAvg}), as functions of the threshold transferred momentum, $Q_\text{th}^2$. The gray areas are accessible with MSSM running when the mixing parameters at production are varied within their current $3\sigma$ bounds (see text), while in the hatched areas only $\theta_{13}$ and the phases are varied. The bands limited by dashed (red) lines are the regions of the ratios accessible with standard oscillations only, i.e., eq.~(\ref{EqProbPabQ}) with $U_\nu^\prime = U_\nu$, when the mixing parameters are allowed to vary in the same manner. In the integral in eq.~(\ref{EqProbAvg}), we have restricted the transferred momentum values to $Q^2 > Q_\text{th}^2$.}
    \label{FigPavgvsQ2}
  \end{center}
\end{figure}

In figure \ref{FigPavgvsQ2} we have plotted the cross-section-averaged probabilities, eq.~(\ref{EqProbAvg}), 
as functions of $Q_\text{th}^2$, fixing the neutrino energy at $E_\nu = 10^{12}$ GeV. As in figure \ref{FigProbvsQ2}, to generate the gray regions, we have allowed the mixing parameters at production to vary within their $3\sigma$ experimental bounds, while the CP and Majorana phases have been varied between $0$ and $2\pi$. The hatched regions were generated by varying only $\theta_{13}$ and the phases, while keeping all of the other mixing parameters at production at their best-fit values. The dashed (red) lines are included for comparison to the standard-oscillation probabilities. In the absence of MSSM running effects, the probabilities would have to be confined between these lines.

From the definition in eq.~(\ref{EqProbAvg}), we see that the average probability receives, at a given value of $Q_\text{th}^2$, the accumulated effects from $Q_\text{th}^2$ up to $\left(2 \times 10^{12}\right) m_N $ GeV$^2$. Thus, if we assumed naively that the values of the weights in the integrand, i.e., of the differential cross section, remain the same throughout the whole range of integration, then deviations of $\langle P_{\alpha\beta}\left(Q_\text{th}^2\right) \rangle^{\overset{\brabar}{\nu}}$ from the no-running expectations would occur for all values of $Q^2_{\text{th}}$. However, in figure \ref{FigPavgvsQ2}, the starting point of the deviations is around $10^{5}$ GeV$^2$. For values of $Q_\text{th}^2 \lesssim 10^5$ GeV$^2$, the accumulated effects are not visible, and hence the probabilities under MSSM running coincide with the corresponding ones under no running. The reason is that the pdf's decrease with growing $x$, which means that the weights in eq.~(\ref{EqProbAvg}) become irrelevant at high values of $Q^2$ compared to those at low $Q^2$. In contrast, for $Q_\text{th}^2 \gtrsim 10^5$ GeV$^2$, the weights evaluated at $Q^2 \sim Q_\text{th}^2$ are comparable to the ones evaluated at slightly higher values of $Q^2$, with the result that the average probabilities start deviating. Finally, note that, with the exception of the differences in the starting points of the deviations, the probability regions in figures \ref{FigProbvsQ2} and \ref{FigPavgvsQ2} have similar shapes and reach the same minimum and maximum values.

The cross-section-averaged flavour fluxes at Earth for the four scenarios, $\langle \Phi_{\overset{\brabar}{\nu}_\alpha}\left(E_\nu\right) \rangle$, are given by eqs.~(\ref{EqFlavRatiosScA})\---(\ref{EqFlavRatiosScD}) after changing $P_{\alpha\beta} \rightarrow \langle P_{\alpha\beta}\left(E_\nu\right) \rangle^{\overset{\brabar}{\nu}}$. These flavour fluxes can be used to define the cross-section-averaged flavour ratios $\langle T \rangle$ and $\langle R \rangle$, respectively, by eqs.~(\ref{EqDefT}) and (\ref{EqDefR}), with the replacement $\Phi_{\overset{\brabar}{\nu}_\alpha} \rightarrow \langle \Phi_{\overset{\brabar}{\nu}_\alpha} \rangle$. 

\begin{figure}[t!]
 \begin{center}
  \scalebox{0.6}{\includegraphics{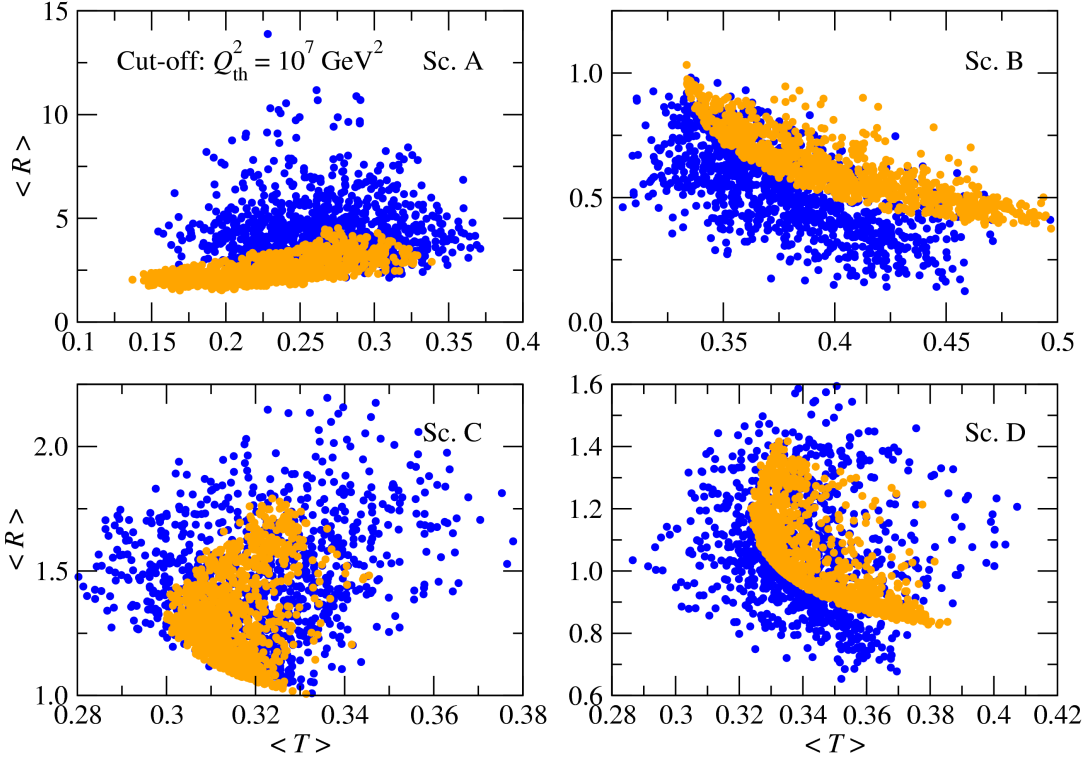}} 
  \caption{Scatter plots of $\langle R \rangle \equiv \langle \tilde{\Phi}_{\nu_e+\bar{\nu}_e} \rangle / \langle \tilde{\Phi}_{\nu_\tau+\bar{\nu}_\tau} \rangle$ vs. $\langle T \rangle \equiv \langle \tilde{\Phi}_{\nu_\mu+\bar{\nu}_\mu} \rangle$ for the four different production scenarios A\---D (see table \ref{TblScenarios}). Orange points were calculated using standard oscillations only, while blue points were calculated under the MSSM running, with an energy $E_\nu = 10^9$ GeV and a cut-off of $Q_\text{th}^2 = 10^7$ GeV$^2$.}
\label{FigRTsemiexp}
 \end{center}
\end{figure}

As expected, we find no difference between the cases of no running and MSSM running for either $\langle R \rangle$ or $\langle T \rangle$ if a low cut-off or no cut-off at all is imposed. Therefore, in order to bring out any differences between the two cases, we show in figure~\ref{FigRTsemiexp} the scatter plot of $\langle R \rangle$ vs.~$\langle T \rangle$, for a fixed energy of $E_\nu = 10^9$ GeV and a cut-off $Q_\text{th}^2 = 10^7$ GeV$^2$. Since this energy is close to the maximum neutrino energy expected, any observed deviations of the probabilities will be close to the largest possibly achievable. In accordance to the results found in Section \ref{sec:theory}, the largest potential deviations occur for scenarios A and B, corresponding, respectively, to pure electron-flavoured and pure muon-flavoured fluxes. In these cases, the stronger deviations are observable in $\langle R \rangle$.   

\subsection{Experimental perspective}

At first sight, the results presented in figure~\ref{FigRTsemiexp} encourage us to look for the MSSM running effects in high-energy astrophysical neutrinos within an experimental context. However, the requirement of using only scattering events with $Q^2 \geq Q_\text{th}^2 = 10^7$ GeV$^2$ reduces dramatically the size of the sample of scattering events that could potentially reveal the presence of MSSM running in a detection experiment. The situation turns out to be even worse if we take into account that at the energy that we have considered, $E_\nu = 10^9$ GeV, the astrophysical neutrino flux is expected to be very low.   

\begin{table}
 \begin{center}
   \small
   \begin{tabular}{|c|c|c|c|c|}
    \hline
    Scenario & \multicolumn{2}{c|}{WB~ $(\times 10^{6})$} & \multicolumn{2}{c|}{BB~ $(\times 10^{7})$} \\
    \cline{2-5}
    & Std. osc. & MSSM & Std. osc. & MSSM \\
    \hline
    A & $0.95\--2.44$ & $1.12\--2.70$ & $1.16\--2.96$ &  $1.34\--3.25$ \\
    \hline
    B & $2.51\--3.80$ & $2.22\--3.74$ & $3.70\--5.60$ & $3.32\--5.51$ \\
    \hline
    C & $1.75\--3.16$ & $1.69\--3.27$ & $2.58\--4.66$ & $2.50\--4.80$ \\
    \hline
    D & $2.01\--3.38$ & $1.87\--3.43$ & $2.96\--4.98$ & $2.77\--5.03$ \\
    \hline
   \end{tabular}
   \caption{\label{TblNumEvts}Ranges of values of the integrated number of $\left(\nu_\mu+\bar{\nu}_\mu\right)$ expected in IceCube-80 (effective volume $V_\text{eff} = 1$ km$^3$) in the range $10^5 \leq E_\nu/\text{GeV} \leq 10^{12}$, after $t= 15$ years of exposure, using standard (no-running) and MSSM flavour transitions, for the WB and BB neutrino flux models described in Appendix \ref{AppendixB}. A cut-off of $Q_\text{th}^2=10^7$ GeV$^2$ has been applied. The ranges were found by varying the mixing parameters at production within their $3\sigma$ bounds and the phases between $0$ and $2\pi$.}
 \end{center}
\end{table}
 
In order to illustrate this situation, we have calculated the minimum and maximum integrated number of muon-neutrinos and anti-neutrinos between $10^5$ and $10^{12}$ GeV, after $15$ years of exposure at IceCube, under two assumptions on the UHE diffuse neutrino flux from AGN, one by Waxman-Bahcall (WB) and the other by Becker-Biermann (BB) (see Appendix \ref{AppendixB}), and applying a cut-off of $Q_\text{th}^2=10^7$ GeV$^2$. The minimum and maximum numbers of events were obtained by varying the mixing parameters at production within their $3\sigma$ bounds and all the phases between $0$ and $2\pi$. These numbers are displayed in table~\ref{TblNumEvts}. There is a strong trade-off between the value of $Q^2_\text{th}$ and the expected number of neutrino events that exhibit a reconstructed $Q^2$ equal to or higher than our threshold: the cut-off we have imposed ($Q^2_\text{th}=10^7$~GeV$^2$) dramatically decreases the number of events to $10^{-6}\--10^{-7}$ in $15$ years, rendering the potential observation of SUSY running through the flavour fluxes null. Thus, detection of a few events would require a \u{C}erenkov detector with an unrealistic effective volume roughly $10^6$ (for the WB flux) to $10^7$ (for the BB flux) times larger than IceCube's. Using the lowest possible SUSY-enhancing cut-off, $Q_\text{th}^2 = 10^6$ GeV$^2$, would increase the event yield in about three orders of magnitude, but still not enough for actual detection. Under these prospects, it makes no sense to calculate the expected reconstructed values of the ratios $R$ and $T$. Therefore, it is evident that IceCube has no actual sensitivity to MSSM running in the flavour ratios: we are hampered both by the $Q^2$ cut-off and by the low flux of UHE neutrinos.  

Although alternative neutrino detection techniques could be considered, such as radio~\cite{Connolly:2008tu} or sound detection~\cite{Abbasi:2009si}, these would still be limited by the low neutrino flux. Furthermore, they would also have to be able to reconstruct the transferred momentum in each interaction and the high $Q^2$ cut-off needed would also reduce the sample size drastically. It seems that presently it is technically unfeasible to detect the effects of MSSM running in the UHE astrophysical neutrino flavour ratios.

\section{Summary and conclusions}
\label{conclusion}

We have analysed the possibility of observing the renormalisation group running of the neutrino mixing parameters in the MSSM. To this end, we have studied the possibility of observing distortions in the extragalactic high-energy neutrino flavour ratios, where the running effects participate through neutrino flavour-transition probabilities. With the inclusion of SUSY effects, modifications to these probabilities with respect to their standard values grow with the transferred momentum in neutrino-nucleon deep inelastic scattering, $Q^2$, which is largest for UHE neutrinos from AGN.

It has been observed that the transition probability under MSSM running starts to differ from the one with no running at $Q^2 = 10^6$ GeV$^2$, which corresponds to our choice for the SUSY scale $\Lambda_\text{SUSY}$. The maximum variation is in the order of $50$\% for $P_{\mu\tau}$. One important feature of these SUSY effects is that they are mainly controlled by the Dirac and Majorana CP phases, and by $\theta_{13}$. We have found that in order to enhance the SUSY effects two conditions must be satisfied: a large value of $\tan\beta$ and a large value for the sum of the neutrino masses. We have set $\tan\beta = 50$ and $\Sigma m_{\nu_i} < 1.3$ eV, following the latest WMAP-only bound. 

We have studied $R = \Phi_{\nu_e+\bar{\nu}_e}/\Phi_{\nu_\tau+\bar{\nu}_\tau}$ and $T = \Phi_{\nu_\mu+\bar{\nu}_\mu}/\left(\Phi_{\nu_e+\bar{\nu}_e}+\Phi_{\nu_\mu+\bar{\nu}_\mu}+\Phi_{\nu_\tau+\bar{\nu}_\tau}\right)$, taking into account the MSSM effects on the oscillation probabilities, for four different production model scenarios, $\left(\Phi_{\nu_e+\bar{\nu}_e}^0:\Phi_{\nu_\mu+\bar{\nu}_\mu}^0,:\Phi_{\nu_\tau+\bar{\nu}_\tau}^0\right)  = \left(1:0:0\right)$, $\left(0:1:0\right)$, $\left(1/2:1/2:0\right)$, and $\left(1/3:2/3:0\right)$, considering the full variation of the neutrino mixing parameters within their current $3\sigma$ allowed ranges.  
We have found that, under MSSM running, for $Q^2 \sim 10^{11}$ GeV$^2$, or even for smaller values, $R$ is able to reach, within the $\left(1:0:0\right)$ production scenario, values 18 times larger than the corresponding ones in the no-running case, while for $T$ the differences are not nearly as significant. In the remaining neutrino production scenarios, the differences between the MSSM running and no-running cases are less pronounced, more so for $\left(1/2:1/2:0\right)$ and $\left(1/3:2/3:0\right)$. 

In order to give a more realistic view of our observations on $R$ and $T$, we have built average forms of these observables using as averaging weights the DIS cross section (i.e., the cross section at detection). We have noted that, in order to obtain differences between the MSSM-running from the standard behaviour in our predictions of $\langle R \rangle$ and $\langle T \rangle$, it is necessary to impose a cut-off condition of $Q^2 \geq Q_\text{th}^2 = 10^7$ GeV$^2$. Otherwise, if we consider lower values of the threshold $Q_\text{th}^2$, the differences completely disappear. This result is discouraging for going forward in the search for MSSM effects in the flavour ratios, since two factors suppress the event rate at a large neutrino telescope: the expected low UHE astrophysical neutrino flux and the fact that the event sample at high $Q^2_\text{th}$ is too low. We have estimated the number of events at IceCube to be $10^{-6}\--10^{-7}$ after 15 years, for a cut-off of $Q_\text{th}^2 = 10^7$ GeV$^2$.

Therefore, although there are corrections to the flavour transition probability due to the MSSM running of the mixing parameters, we are compelled to conclude that it is not feasible, with the available and foreseeable technology, to detect the presence of such effects in the flavour ratios of UHE astrophysical neutrinos.

\acknowledgments

The work of M.~B. and A.~M.~G. was supported by the Vicerrectorado de Investigaci\'on at PUCP through Grant No. DGI-2010-0021. J.~J.~P. would like to thank Takashi Shimomura, Oscar Vives, Jos\'e Bernabeu and particularly Arcadi Santamar\'ia for fruitful discussions. He would also like to thank the Pontificia Universidad Cat\'olica del Per\'u (PUCP) for its warm hospitality during his visit. The work of J.~J.~P. was partially supported by the Spanish MICINN and FEDER (EC) Grant No. FPA2008-02878. This work was made possible by the partial support of ALFA-EC funds in the framework of the HELEN Project.

\appendix

\section{Neutrino flavour-transition probability with different mixing matrices}\label{AppendixA}

The probability of a neutrino produced with flavour $\alpha$ interacting with flavour $\beta$ can be expressed as:
\begin{equation}\label{EqProbG1}
 P_{\alpha\beta} = \sum_{i=1}^3 g_{\alpha i} g_{\beta i}^\prime ~,
\end{equation}
where the coefficients $g_{\alpha i}$ depend only on $\left(\theta_{12},\theta_{13},\theta_{23},\delta\right)$ and the $g_{\beta i}^\prime$ depend only on $\left(\zeta_{12},\zeta_{13},\zeta_{23},\delta_1\right)$ and, through the RGEs, on $Q$. Their expressions are shown in table~\ref{TblCoefficients}. Thus, we see that the probability can be written as a sum of terms, each of which is the product of a function of only standard mixing parameters times a function of only renormalized mixing parameters.

Note that the standard probabilities can be recovered simply by taking $\zeta_{ij} \rightarrow \theta_{ij}$ and $\delta_1 \rightarrow \delta$, which is equivalent to the replacement $g_{\beta i}^\prime \rightarrow g_{\beta i}$ in eq.~(\ref{EqProbG1}).

\begin{table}
 \begin{center}
   \begin{tabular}{|c|c|c|c|}
    \hline
    Coefficient name & Expression                                                             & Best-fit value       \\
    \hline
    $g_{e1}$         & $c_{\theta_{13}}^2 c_{\theta_{12}}^2$                                  & $0.67$               \\
    $g_{e2}$         & $c_{\theta_{13}}^2 s_{\theta_{12}}^2$                                  & $0.32$               \\
    $g_{e3}$         & $s_{\theta_{13}}^2$                                                    & $0.01$               \\
    $g_{\mu1}$       & $f\left(\theta_{12},\theta_{13},\theta_{23},\delta\right)$             & $0.17+0.045c_\delta$ \\
    $g_{\mu2}$       & $f\left(\theta_{12}+\pi/2,\theta_{13},\theta_{23},\delta\right)$       & $0.37-0.045c_\delta$ \\
    $g_{\mu3}$       & $c_{\theta_{13}}^2 s_{\theta_{23}}^2$                                  & $0.46$               \\
    $g_{\tau1}$      & $f\left(\theta_{12}+\pi,\theta_{13},\theta_{23}+\pi/2,\delta\right)$   & $0.15-0.045c_\delta$ \\
    $g_{\tau2}$      & $f\left(\theta_{12}+\pi/2,\theta_{13},\theta_{23}+\pi/2,\delta\right)$ & $0.32+0.045c_\delta$ \\
    $g_{\tau3}$      & $c_{\theta_{13}}^2 c_{\theta_{23}}^2$                                  & $0.53$               \\
    \hline
   \end{tabular}
 \end{center}
\caption{\label{TblCoefficients}$g$-functions used to calculate the flavour-transition probability, eq.~(\ref{EqProbG1}). The expressions for $g_{\alpha i}^\prime$ are obtained from the corresponding expressions for $g_{\alpha i}$ shown here, after the replacements $\theta_{ij} \rightarrow \zeta_{ij}$ and $\delta \rightarrow \delta_1$. The auxiliary function $f$ is defined as $f\left(\theta_1,\theta_2,\theta_3,\delta\right) \equiv c_{\theta_3}^2 s_{\theta_1}^2 + \frac{1}{2} c_\delta s_{2\theta_1} s_{2\theta_3} s_{\theta_2} + c_{\theta_1}^2 s_{\theta_2}^2 s_{\theta_3}^2$.}
\end{table}

The standard coefficients $(g_{\alpha i}-g_{\alpha j})$ shown in eqs.~(\ref{EqPbetae})-(\ref{EqPbetatau}) can then be written as:
\begin{eqnarray}
 g_{e1} - g_{e2} 
 &=& 
 c_{\theta_{13}}^2 c_{2\theta_{12}} \\
 g_{e1} - g_{e3}
 &=& 
 c_{2\theta_{13}} - s_{\theta_{12}}^2 c_{\theta_{13}}^2 \\
 g_{e3} - g_{e2}
 &=& 
 -c_{2\theta_{13}} + c_{\theta_{12}}^2 c_{\theta_{13}}^2 \\
 g_{\mu1} - g_{\mu2}
 &=& 
 c_\delta s_{2\theta_{12}} s_{2\theta_{23}} s_{\theta_{13}}
 - c_{2\theta_{12}} \left( c_{2\theta_{23}} + c_{\theta_{13}}^2 s_{\theta_{23}}^2 \right) \\
 g_{\mu1} - g_{\mu3}
 &=&
 \frac{1}{2} c_\delta s_{2\theta_{12}} s_{2\theta_{23}} s_{\theta_{13}}
 + c_{\theta_{23}}^2 s_{\theta_{12}}^2 - \left( c_{2\theta_{13}} + s_{\theta_{12}}^2 s_{\theta_{13}}^2 \right) s_{\theta_{23}}^2 \\
 g_{\mu3} - g_{\mu2}
 &=&
 \frac{1}{2} c_\delta s_{2\theta_{12}} s_{2\theta_{23}} s_{\theta_{13}} 
 -c_{\theta_{23}}^2 c_{\theta_{12}}^2 + \left( c_{2\theta_{13}} + c_{\theta_{12}}^2 s_{\theta_{13}}^2 \right) s_{\theta_{23}}^2 \\
 g_{\tau1} - g_{\tau2}
 &=& 
 -c_\delta s_{2\theta_{12}} s_{2\theta_{23}} s_{\theta_{13}}
 + c_{2\theta_{12}} \left( c_{2\theta_{23}} - c_{\theta_{13}}^2 c_{\theta_{23}}^2 \right) \\
 g_{\tau1} - g_{\tau3}
 &=&
 -\frac{1}{2} c_\delta s_{2\theta_{12}} s_{2\theta_{23}} s_{\theta_{13}}
 + s_{\theta_{23}}^2 s_{\theta_{12}}^2 - \left( c_{2\theta_{13}} + c_{\theta_{12}}^2 s_{\theta_{13}}^2 \right) c_{\theta_{23}}^2 \\
 g_{\tau3} - g_{\tau2}
 &=&
 \frac{1}{2} c_\delta s_{2\theta_{12}} s_{2\theta_{23}} s_{\theta_{13}} 
 - s_{\theta_{23}}^2 c_{\theta_{12}}^2 + \left( c_{2\theta_{13}} + c_{\theta_{12}}^2 s_{\theta_{13}}^2 \right) c_{\theta_{23}}^2
\end{eqnarray}

\section{Number of neutrinos expected at IceCube}\label{AppendixB}

Adapting the expressions from \cite{Anchordoqui:2005gj}, the number of CC interactions initiated by a downgoing\footnote{Since at energies $E_\nu \gtrsim 10^7$ GeV energy losses of Earth-traversing neutrinos in NC $\overset{\brabar}{\nu} N$ interactions become important, i.e., the shadow factor \cite{Gandhi:1995tf} becomes $\lesssim 0.1$ and the Earth turns opaque for upgoing neutrinos, we have calculated only the number of downgoing neutrino events.} astrophysical flux of $\alpha$\---flavoured (anti-)neutrinos in a detector of effective volume $V_\text{eff} = 1$ km$^3$ and opening solid angle $\Omega = 5.736$ sr (corresponding to a zenith angle of up to 85$^\circ$) can be calculated as
\begin{equation}
 N_{\overset{\brabar}{\nu}_\alpha}^\text{CC} 
 = t\, n_T V_\text{eff} \Omega 
     \int_{E_\nu^{\min}}^{E_\nu^{\max}} dE_\nu~
     \sigma_\text{CC}^{\overset{\brabar}{\nu}}\left(E_\nu\right) 
     \langle \tilde{\Phi}_{\overset{\brabar}{\nu}_\alpha}\left(E_\nu\right) \rangle_\text{CC}~
     \Phi_{\nu_\text{all}}\left(E_\nu\right) ~,  \label{EqNEvtsCC}
\end{equation}
where $E_\nu^{\min} = 10^5$ GeV and $E_\nu^{\max} = 10^{12}$ GeV are, respectively, the minimum and maximum neutrino energies that we have considered, $t=15$ years is the exposure time, $n_T = 5.1557 \times 10^{23}$ cm$^{-3}$ is the number density of targets (nucleons) in ice, and $\sigma_\text{CC}^{\overset{\brabar}{\nu}}$ is the \mbox{(anti-)neutrino} CC scattering cross section off an isoscalar target, eq.~(\ref{EqSigmaTot}).

The flux $\Phi_{\nu_\text{all}}$ is the all-flavour diffuse flux of astrophysical neutrinos, for which we have used two models: a conservative one, by Waxman and Bahcall (WB) \cite{Waxman:2002wp} and an optimistic one, by Becker and Biermann\footnote{The BB flux featured here makes use of an updated normalisation \cite{Arguelles:2010yj} calculated using the latest results reported by the Pierre Auger Collaboration on the correlation between the arrival directions of UHE cosmic rays and the positions of know AGN \cite{:2010zzj}. The all-flavour BB flux is estimated from the muon-neutrino BB flux by assuming the standard-oscillation scenario, where $\left(\tilde{\Phi}_{\nu_e+\bar{\nu}_e}:\tilde{\Phi}_{\nu_\mu+\bar{\nu}_\mu}:\tilde{\Phi}_{\nu_\tau+\bar{\nu}_\tau}\right) = \left(1/3:1/3:1/3\right)$, and multiplying any of the individual flavour fluxes by $3$.} (BB) \cite{Becker:2008nf}: 
\begin{eqnarray}
 \Phi_{\nu_\text{all}}^\text{WB}\left(E_\nu\right) 
 &=& 10^{-8} \left(E_\nu/\text{GeV}\right)^{-2} ~\text{GeV}^{-1} ~\text{cm}^{-2} ~\text{s}^{-1} ~\text{sr}^{-1} \\
 \Phi_{\nu_\text{all}}^\text{BB}\left(E_\nu\right) 
 &\simeq& 5 \times 10^{-2} \left(E_\nu/\text{GeV}\right)^{-2.9}
 ~\text{GeV}^{-1} ~\text{cm}^{-2} ~\text{s}^{-1} ~\text{sr}^{-1} ~.
\end{eqnarray}
Both fluxes lie below the upper bound curves reported by different neutrino detection experiments: AMANDA-II \cite{Abbasi:2010ak,Ishihara:2010}, ANITA-II \cite{Gorham:2010kv}, Auger \cite{Abraham:2009uy}, RICE \cite{Kravchenko:2006qc}, IceCube 22-strings \cite{Abbasi:2010ak}, and a preliminary IceCube 40-strings bound \cite{Yoshida:2010,Montaruli:2010hr}. In particular, the BB model parameters were chosen to yield the maximum allowed flux within this model\footnote{The flux $\Phi_{\nu_\text{all}}^\text{BB}$ is calculated using a value for the spectral index of $\alpha = 2.9$, and BB model parameters $\Gamma_\nu/\Gamma_\text{CR} = 2.05$ and $z_\text{CR}^{\max} = 0.019$ (see \cite{Becker:2008nf} for an explanation of the parameters).}.

\end{document}